# On proximity versus geo-information systems

Dmitry Namiot[1], Manfred Sneps-Sneppe[2]

[1]Lomonosov Moscow State University
Faculty of Computational Mathematics and Cybernetics
Moscow, Russia
{dnamiot@gmail.com}

[2]Ventspils University of Applied Science
Ventspils International Radioastronomy Centre
Ventspils, Latvia,
{manfreds.sneps@gmail.com }

**Abstract.** In this paper, we propose the replacement of widely used models of geo-information systems with a new conception based on network proximity. Geo-information systems have attracted great attention and demonstrated big progress in recent times, especially for mobile services. We can point out many objective reasons for this. On the one hand, users require services, mainly at their location, and on the other hand, location determination has become easy, especially due to the proliferation of smartphones. But at the same time, the actual geo-calculation are not needed by most of these services. For the majority of geo-services, geo-coordinates are used only for searching and organizing data. And the meaning of the service is to search for information tied to the current location of the requesting party. In other words, in most cases, service refers to data near the current location. So, our idea is to build services directly on assessing to proximity and completely bypass geo-calculations. It also opens the way for completely new services that were impossible or difficult to implement with geo-computations.

## 1 Introduction

Geo-information systems have attracted attention and have received great development in recent times [1]. The reasons for this are fairly obvious. On the one hand, services (services, information) are required by users, primarily at their location (current, planned), and on the other hand, location determination has become easy, especially in connection with the proliferation of mobile devices (smartphones). But at the same time, the actual geographic coordinates are not needed by the majority of such services. Coordinates are used only for searching and organizing data. And the meaning of the service (service) is to search for information tied to the current location of the requester. In other words, in most cases, service is meant near the current location. Hence the idea to build services directly on the assessment of proximity, completely bypassing the work with coordinates. It also makes possible completely new services, especially in the technical field (for example, in transport, agriculture, etc.).

By and large, all mobile applications should provide (in fact, most of them are exactly so arranged) so-called context-aware (sometimes, context-sensitive) services [2]. A context is any information (measurable data) that can be added to a location. For example, the context may be data of mobile phone sensors, publication in social networks, etc. A context-aware service is a service whose work depends on the context. In particular, we devoted a lot of time to context-aware applications in our previous papers [3, 4]. In this description, an important point for us is that the context, in fact, always includes location. Thus, in the case of mobile devices (mobile services), context-aware services are also geo-informational.

The next element we would like to point attention to is the geographic component of the geo-information services. On the one hand, everything is simple here. There is the World Geodetic System (WGS) standard, the latest revision of which is WGS 84 and is used everywhere. Geographic coordinates (latitude, longitude) can be easily obtained on a mobile device and used in the service (for obtaining the service). On the other hand, in practical applications, are there really so many services that really need geo-coordinates? In fact, most of the so-called geo-information services are designed to provide information / services to subscribers who are in some geographically defined (restricted) area. Or, in other words, for those who are at a limited distance from the center of a given area (not further than some specific distance). That, in turn, can be formulated as the provision of services in the vicinity of this very central point. In other words, the service is actually based on the proximity. And the proximity is understood in this case as a distance metric.

The essential moment here is the fact that geo-coordinates are not at all important for receiving a service. These coordinates are used only to calculate the metric. Accordingly, when describing these most provided services, geographical coordinates are used (in most cases) simply as keys for organizing a search. The real semantics of the most "geo-information" services is to find a service near the current location. This is where the idea of information systems, directly built on data about proximity, arises. Proximity information systems do not even use geographic coordinates as keys, but directly link the provision of information (services), interface structure, etc. to proximity-related information.

Here we can immediately ask a question - there are already existed Proximity Services (ProSe), proposed in the 3GPP specification [5]. Formally, by name, they should fall into this category, but, in fact, it is somewhat different. First introduced in release 12 of the 3GPP specifications, ProSe (Proximity Services) is D2D technology (device-to-device), which allows LTE devices (5G) to detect each other and communicate directly. The key point here is the organization of communication in a device to device pair. From this point of view, the peer to peer network can also be attributed to proximity services. It also connects devices. In our case, we use the same term Proximity Services (due to the fact that these are really services based on a physically close location), but the main point is the lack of orientation towards the interaction (connection) of two devices. This is by no means D2D services. It is about the content rather than connectivity. Also, proximity is considered here not only in relation to two mobile devices (in practice, smartphones). If we consider, for example, the geo-information services with which we began our consideration, it is obvious that, for example, searching for nearby ATMs (a typical geo-information service) does not imply (in most cases) connections with the found devices. A service can report, for example, about other people in the vicinity (there is no device at all to connect with). This is how proximity-based services will be treated in this paper. Typical services (attached to mobile devices) can:

- open (close) any information for users who are close to another device (object, things). Note that transferring data from one device to another nearby is a very special case. For example, there may be no other device for connecting and transferring data. The data itself may be on a third-party device (server), and their availability will be determined by the proximity of the devices. Just a practical example: coupons for discounts for some cafe become available for downloading from a cloud server for mobile subscribers nearby this cafe. There is no connectivity at all, just some data become available for users in proximity only;

- perform any actions after being in the vicinity of the objects (device, things) mentioned above. As well as to bind such an action to the fact of occurrence in the vicinity of some object (device) or, symmetrically, in case of violation of the proximity condition that previously existed. We can draw some parallels with smart contracts [6]. Actions are performed automatically upon the occurrence of conditions associated with proximity;

- change the interface, allow (deny) any possibilities, depending on the occurrence (termination) of proximity conditions.

The first two points refer to the so-called context-aware systems. The latter - to adaptive interfaces (in English literature - ambient intelligence (AMI) [7])

A simple question may arise - what could be the reasons, nevertheless, not to use geographic coordinates? Here are the following considerations:

- The energy intensity of the process. Getting a position is simple, but rather expensive for a battery of a mobile device, if it is done frequently. In fact, it explains all sorts of combined methods (such as assisted GPS, for example [8]) for positioning.

- Indoor positioning. GPS query may not be possible.

- Positioning accuracy. It is quite possible to get greater positioning accuracy without GPS

- GPS signal may be muffled / changed

- Access to a service based only on geographic coordinates is more difficult to restrict (a location request is available to all)

One of the most important issues: geographical coordinates do not change. And if we base the service on proximity to objects (device, things), which itself moves, then we get a completely new class of services, which was impossible with geographic coordinates. The service (available information, possible actions, etc.) is tied to the current location of the reference object and moves with it. By analogy with the geo-fence (geo-grid), we can talk about proximity-fence.

The rest of the paper is organized as follows. In Section 2, we discuss related works. In Section 3, we discuss proximity measurements and the importance of network proximity. In Section 4, we present proximity information systems examples.

## 2 Related works

What is usually considered in the literature (and practical projects) in relation to geo-information services?

The structure of any typical geo-information service is as follows:

- there is a database (storage) of information objects with the well-known geographic locations. For example, in the simplest case - the point of interests and their geographical coordinates. In other words, there are some geographical places (real or virtual) with their descriptions. Instead of a geographic location, there may be a description of some service, events, etc. Instead of a pair of coordinates (latitude, longitude), there may be information about a certain geographic area (geo-lattice), which is, for example, a description of the range, service area, etc. [9];

- the geographical coordinates of the user (users) are determined. For example, via GPS services on mobile devices or via some combined methods (mobile operators, assisted GPS positioning, etc.) [10];

- the engine calculates distances (metrics) for the user's location and possible geographical positions, zones, etc. [9];

Accordingly, the basic issues that should be addressed in such systems are as follows:

- an efficient storage organization. Here we can note, for example, geo-extensions for databases;

- an organization of effective data collection;

- an organization of effective queries related to geo-calculations. For example, calculating distances, checking whether a point belongs to a certain area, searching for intersections for areas;

- application program interfaces (API). Most often, geo-informational solutions do not exist by themselves, but are part of other software systems. Accordingly, for their integration, we need programmatic access for the above-described features.

In terms of working with content, these are the above-mentioned context-aware systems and so-called Ambient Intelligence (AMI) [11]. We note once again that proximity is in no way connected with direct interactions (D2D). This is just a metric relating to the physical location of two or more objects.

## 3. How to measure a proximity

First, it is necessary to mention the sensors. The proximity sensor on a mobile device allows you to determine the approximation of an object (object) without physical contact with it. For example, a simple infrared proximity sensor has a working measurement distance of several tens of centimeters and a detection angle of several tens of degrees. Unmanned vehicles are equipped with both radar and lidar, which, in fact, are designed to identify nearby objects (obstacles). Image processing for a camera on a mobile device can give information about the proximity. Alternatively, we can try to find some common characteristics (measurements) for two mobile devices. For example, try to analyze the sound in the vicinity of each of the mobile devices and compare the measurement characteristics among themselves, etc.

Common problems that can be noted here:

- there is no general model. For example, installed sensor sets vary for different phone models;

- there is no common (unified) API for such measurements;

- some methods of determining proximity may be practically unsuitable (for example, require a large number of actions from mobile users, impose restrictions on other programs, etc.). For example, getting images just for checking the proximity is too disruptive.

From this explanation, it follows that from a practical point of view, the most important is the use of network proximity. This refers to the assessment of the proximity of devices using network technologies (in the first place - wireless networks). By definition, as it was historically introduced, a smartphone is a mobile phone with network capabilities. Wireless protocols (for example, Wi-Fi, Bluetooth) have a limited range. The availability of, for example, a Wi-Fi access point already gives information about the proximity of the mobile device to this access point. Mobile device maintenance of a particular cell tower is also proximity information: two devices served by one tower are obviously in some proximity. By definition, the signal path for wireless networks is limited. Accordingly, the fact of receiving a signal indicates some proximity to its source. Transitively, two devices (for example, mobile phones) in the vicinity of the signal source are located close to each other.

Thus, to determine the network proximity, we have the signal itself, the reception of which is confirmed by the availability (reading) of a certain value (for example, the Wi-Fi access point name, the Bluetooth node address, and so on), as well as a number of measurements, among which the most frequently used is the relative signal strength (RSSI). This allows you to consider (interpret) network interfaces as sensors. At the same time, the operating systems of mobile devices provide the background operation of such "sensors" without conflicts with other applications. The software interface is supported by the operating system of the mobile device and, thus, is unified.

So, the network proximity is the main tool for proximity measurements (at least, for now) [12]. Note that wireless networks, for example, are already used in navigation (positioning) tasks. For example, knowing the geographical coordinates of Wi-Fi access points and measured signal characteristics, we can estimate (with varying accuracy) the distance to them and the geographical location of the mobile device. This is often used to refine positioning (as an addition to GPS). But there are some limitations, which have already been mentioned above. This scheme can work when the picture of the network is static. The reference points are known, their coordinates are measured and, in most cases, the preparation of the radio map is done. But in this paper, we consider systems where a determination of coordinates and geo-calculations are excluded. Accordingly, a proximity will be defined as a proximity (by distance) to network elements or other mobile devices but calculated solely by means of network technologies.

## 4. On proximity-based information systems

In this section, we would like to highlight specific examples of information systems based on the use of the concept of proximity. This section summarizes our experience in this kind of implementation and is an attempt to build a classification for this class of systems.

1. Providing access to some data, depending on the nearby wireless networks. Options and examples of use:

*in B2C version*: some of the elements of the manufacturer's (seller's, provider's) website become available (visible) to the mobile user when this user is located close to the location of service provision. Proximity is determined by the visibility (availability) of a Wi-Fi network or a specially created Bluetooth node (Bluetooth tags);

*in B2B version*: the identification of the object determines the description of the provided service, available services, prices, etc. Note that here a similar model can be implemented using the Google Physical Web. In this case, the tags send some URL that can be used to access the information. Also here you can draw parallels with the QR-which is present on the physical object. However, unlike QR codes, visual scanning is not necessary. Access to the identification of wireless objects is carried out programmatically, and such a model is naturally more adapted to machine-to-machine interaction (M2M). As a special case, we can note the implementation of such systems on the moving objects. For example, the Bluetooth point associated with the multimedia panel of the car can be used as a program-accessible (program-readable) identifier of this car, which can be used in service applications. It is obvious that such an ID will be available for vehicles in the vicinity of the decision point. By a similar manner it will work with wireless access nodes in public transport;

*in C2C option*: a mobile subscriber opens a wireless network point directly on his phone and publishes his own information associated with this point. For example, in the Android operating system, you can programmatically create a Bluetooth access point. In this case, one mobile phone will be enough to publish the data and present it to other mobile subscribers nearby. The data will "move" along with the movement of the publisher's phone. Typical use is for classifieds.

As per examples for such kind of systems, we could mention, for example, SpotEx (Spot Expert) system [13] and Car as tags model [14]. They are based on the production system (set of rules) determining the visibility (availability) of information on the availability of wireless networks. The condition for the rules is some fingerprint of the web. The fingerprint here is a list of network nodes with appropriate limits of RSSI values. In other words, a typical condition for production rule looks so:

   IF wireless_node_with_address (M) IS VISIBLE and RSSI IS BETWEEN [A, B] THEN *Perform some action*

   The advanced model includes tools for describing proximity conditions in terms of fuzzy logic using FCL [16].

At the same time, data in such systems can be entered both centrally and by the users themselves (user-generated content).

2. Providing a physical expansion of social networks. The point is that there may be different variations of the service, which shows the users of the social network (s) nearby. The general scheme of implementation is as follows. The user is authorized in the social network (Facebook, Linkedin), receives an ID (unique user ID in the social network), forms a URL with this ID and distributes it as an identifier of the Bluetooth access point opened right on the own phone. As a result, other users of the same application will be able to see the identifiers of available access points (user identifiers), which lead to social network accounts. This is a substitute for check-in (marks on a presence in the social network). Note that it is not necessary to create any entries in the social network, the social network (its API) is used only to confirm the user's identification.

3. Mesh network model for information transfer. An example is Bluetooth radio [15]. Information that one user transmits (translates by changing the identification in the wireless network) is automatically, when it is received by another user, it is automatically relayed by it. This is an analogue of data distribution in the mesh network. This is a mechanism that can actually be used to create a mobile ad hoc network.

## 5. Conclusion

In this article, we presented the foundations, possible and existing solutions for information systems that are based on the concept of proximity. In our opinion, this approach is quite relevant and popular addition (and in many cases - direct replacement) for geo-information systems. Systems based on proximity in many cases have their distinct advantages over traditional geo-information systems. The article discusses in detail the possible application models, as well as their main advantages and reasons for their use. And the proliferation of smartphones makes the implementation of systems based on network proximity (the main tool for measuring proximity today) simple and cheap.

## 6. Acknowledgement

We would like to thank the staff of the Open Information Technologies Laboratory of the Faculty of the CMC of the Lomonosov Moscow State University for useful discussions, criticism, and assistance in working on software implementations.